\documentstyle[12pt]{article}

   \font\tenmsb=msbm10 scaled\magstep 1
   \font\sevenmsb=msbm7 scaled \magstep 1
   \font\faivemsb=msbm5 scaled \magstep 1
\newfam\msbfam
      \textfont\msbfam=\tenmsb
      \scriptfont\msbfam=\sevenmsb
      \scriptscriptfont\msbfam=\faivemsb

\font\tengothic=eufm10 scaled\magstep 1
\font\sevengothic=eufm7 scaled\magstep 1
\newfam\gothicfam
      \textfont\gothicfam=\tengothic
      \scriptfont\gothicfam=\sevengothic

\newcommand{\be}{\begin{equation}}
\newcommand{\ee}{\end{equation}}

\newcommand{\dlt}{\delta}

\newcommand{\bt}{\beta}
\newcommand{\vp}{\varphi}
\newcommand{\ep}{\varepsilon}
\newcommand{\al}{\alpha}
\newcommand{\ra}{\rightarrow}
\newcommand{\sgm}{\sigma}

\newcommand{\gm}{\gamma}
\newcommand{\om}{\omega}

\begin{document}

\begin{center}

{\Large{\bf Calculation of critical exponents by self-similar 
factor approximants} \\ [5mm]
V.I. Yukalov$^{1,2}$ and E.P. Yukalova$^3$} \\ [3mm]

{\it $^1$Bogolubov Laboratory of Theoretical Physics, \\
Joint Institute for Nuclear Research, Dubna 141980, Russia, \\ [2mm]

$^2$Department of Technology and Economics, \\
Swiss Federal Institute of Technology, Z\"urich CH-8032, Switzerland, \\ 
[2mm] 

$^3$Laboratory of Information Technologies, \\
Joint Institute for Nuclear Research, Dubna 141980, Russia}

\end{center}

\vskip 3cm

\begin{abstract}

The method of self-similar factor approximants is applied to calculating
the critical exponents of the $O(N)$-symmetric $\vp^4$ theory and of the 
Ising glass. It is demonstrated that this method, being much simpler than 
other known techniques of series summation in calculating the critical 
exponents, at the same time, yields the results that are in very good 
agreement with those of other rather complicated numerical methods. The 
principal advantage of the method of self-similar factor approximants is 
the combination of its extraordinary simplicity and high accuracy.

\end{abstract}

\vskip 1cm

{\bf PACS}: 

05.70.Jk Critical point phenomena in thermodynamics,

02.30.Lt Sequences, series, and summability,

02.30.Mv Approximations and expansions

\section{Introduction}

The knowledge of critical exponents, characterizing critical phenomena, 
provides us with basic information on the behavior of thermodynamic 
quantities in the vicinity of critical points [1--3]. This is why so much 
efforts have been devoted to the experimental measurements as well as to 
theoretical calculations of these exponents.

Because of the complexity of realistic theoretical models, critical 
exponents can usually be represented by power series obtained with the help 
of some perturbation theory. Such series are practically always divergent, 
which requires to use resummation techniques allowing for the determination 
of effective limits of divergent series. The standard approach applied to 
the summation of series, associated with critical indices, is based on the 
Pad\'e-Borel-Leroy transformation and its variants [4]. Another approach is 
based on optimized perturbation theory [5], where the resummation is due to 
control functions transforming divergent series into converging ones. 
Introducing control functions through the change of expansion variables, it 
is possible to resum the field-theoretic expansions for critical exponents 
[6,7]. The convergence of the optimized perturbation theory can be greatly 
accelerated by invoking the self-similar approximation theory [8--16], as has 
been done for calculating critical exponents [17]. All these approaches, 
mentioned above, require quite complicated numerical calculations. A purely 
numerical procedure of calculating critical exponents is due to Monte Carlo 
simulations [18--29].

In the present paper, we suggest a new approach for the summation of series 
related to critical exponents. This approach uses the method of self-similar 
factor approximants [30--33], whose mathematical foundation is based on the 
self-similar approximation theory [8--16]. The construction of the factor 
approximants is very simple and straightforward. We recall the main 
definitions in Section 2. Then, in Section 3, we apply these approximants 
for the summation of the $\ep$-expansions for the critical exponents of the 
$N$-vector $\vp^4$ field theory. Our very simple method yields the results 
that are in perfect agreement with the most complicated numerical procedures. 
In Section 4, we demonstrate that the suggested method is applicable even for 
such a notoriously difficult problem as finding the critical indices from 
the high-temperature series expansions for spin glasses. Finally, Section 5 
is conclusion.

\section{Self-similar factor approximants}

Suppose that our aim is to reconstruct a real function $f(x)$ of a real 
variable $x$, when the function is represented by its asymptotic expansion 
at $x\ra 0$ as a power series
\be
\label{1}
f_k(x) = \sum_{n=0}^k a_n x^n \; ,
\ee
where $k=0,1,2,\ldots$. Without the loss of generality, we may assume that 
$a_0=1$ in expansion (1). This is because if instead of form (1) we would have
a more general expression
$$
f^{(k)} = f^{(0)}(x) \sum_{n=0}^k a_n' x^n \; ,
$$
with a given function $f^{(0)}(x)$, then we could return to expansion (1), 
with $a_0=1$, by defining
$$
f_k(x) \equiv \frac{f^{(k)}(x)}{f^{(0)}(x)\; a_0'}  \; .
$$
The sequence $\{f_k(x) \}$ is usually divergent for any finite $x$.

The method of self-similar factor approximants [30--33] makes it possible to 
extrapolate the asymptotic expansion (1), valid only for $x\ra 0$, to the whole 
region of the variable $x\geq 0$. When $k=2p$ is even, with $p$ being an integer, 
then the even-order factor approximant is
\be
\label{2}
f_{2p}^*(x) =\prod_{i=1}^p \left ( 1 + A_i x\right )^{n_i} \; .
\ee
The parameters $A_i$ and $n_i$ are obtained from the re-expansion procedure, 
that is, by expanding approximant (2) in powers of $x$ up to the $k$-th order 
and equating the latter expansion with the initial one given by form (1). 
This re-expansion procedure yields the set of $2p$ equations
\be
\label{3}
\sum_{i=1}^p n_i A_i^n = B_n \qquad (n=1,2,\ldots,2p)
\ee
for $2p$ parameters $n_i$ and $A_i$, with the right-hand sides
$$
B_n \equiv \frac{(-1)^{n-1}}{(n-1)!} \; \lim_{x\ra 0}\; \frac{d^n}{dx^n}\;
\ln f_k(x) \; .
$$
As is evident, the quantities $n_i$, $A_i$, as well as $B_n$, depend on the 
considered order $k$. But for avoiding excessively cumbersome notations, we do 
not use here the double labelling. When $k=2p+1$ is odd, the odd-term factor 
approximant is
\be
\label{4}
f_{2p+1}^*(x) =\prod_{i=1}^{p+1} \left ( 1 + A_i x\right )^{n_i} \; ,
\ee
with the parameters $A_i$ and $n_i$ defined by the re-expansion procedure 
yielding the set of equations
\be
\label{5}
\sum_{i=1}^{p+1} n_i A_i^n = B_n \qquad (n=1,2,\ldots,2p+1) \; ,
\ee
with the scaling condition $A_1=1$.

In this way, for any given expansion (1), the construction of self-similar 
factor approximants (2) or (4) is rather simple and straightforward. It has 
been shown [30--33] that the factor approximants are more general and accurate 
than Pad\'e approximants, having, in addition, a principal advantage of being 
uniquely defined. This means that for each given order $k$ of expansion (1) 
there is just the sole factor approximant, while for each $k$ there exists 
a table of $k$ different Pad\'e approximants $P_{[M/N]}$, with $M+N=k$. There 
is no general recipe for choosing one of the $k$ available Pad\'e approximants. 
One often chooses the diagonal ones, but, as is easy to show, the latter are 
not always the most accurate ones. Such a problem of multiple possibilities 
does not arise for factor approximants: for each $k$, of the expansion $f_k(x)$, 
there is just one factor approximant $f_k^*(x)$.

\section{Exponents for $O(N)$-symmetric theory}

Let us consider the $N$-component vector $\vp^4$ field theory, for which the
critical exponents can be obtained in the form of the $\ep$-expansions, with 
$\ep\equiv 4-d$, and $d$ being the space dimensionality. The derivation of 
these dimensional expansions can be found in the book [6]. In the Appendix A, 
we give the expansions that are considered in the present section. As is known, 
such expansions are divergent and require a resummation procedure. To illustrate 
more explicitly how the method of self-similar factor approximants works, let 
us start with the scalar single-component field $(N=1)$. Then we have
$$
\eta \simeq 0.0185\ep^2 + 0.0187\ep^3 - 0.0083\ep^4 + 0.0257 \ep^5 \; ,
$$
$$
\nu^{-1} \simeq 2 - 0.333\ep - 0.117\ep^2 + 0.124\ep^3 - 0.307 \ep^4 + 
0.951\ep^5 \; ,
$$
\be
\label{6}
\om \simeq \ep - 0.63\ep^2 + 1.62\ep^3 - 5.24 \ep^4 + 20.75\ep^5 \; .
\ee
We reduce each of these expansions to the form
\be
\label{7}
f_k(\ep) = f_0(\ep) \sum_{n=0}^k a_n \ep^n \; ,
\ee
where $a_0=1$. According to Section 2, we construct the factor approximants
\be
\label{8}
f_k^*(\ep) = f_0(\ep) \prod_{i=1}^{N_k} \left ( 1 + A_i\ep
\right )^{n_i} \; ,
\ee
in which
\begin{eqnarray}
\nonumber
N_k = \left \{ \begin{array}{ll}
\frac{k}{2}\; , & k=2p= 2,4,\ldots \\
\frac{k+1}{2}\; , & k=2p+1 = 3,5,\ldots
\end{array} \right.
\end{eqnarray}
and the parameters $A_i$ and $n_i$ are obtained from the re-expansion procedure.
Setting $\ep=1$, we find the desired approximation $f_k^*(1)\equiv f_k^*$. The 
error bar for the approximant $f_k^*$ is given by
$$
\pm \; \frac{1}{2}\left ( f_k^* - f_{k-1}^* \right ) \; \qquad k=2,3,\ldots
$$

Reducing the series for $\eta$ to form (7), we have $\eta_0(\ep)=0.0185\ep^2$.
Constructing the factor approximants (8), we find for $\eta_2^*(\ep)$ the 
parameters
$$
A_1 = 1.898511\; , \qquad  n_1 = 0.532423\; ,
$$
and for $\eta_3^*(\ep)$, we get
$$
A_1 = 1\; , \qquad n_1 = 0.789878 \; , \qquad A_2 = 5.110862\; ,
\qquad n_2 = 0.043228\; .
$$
Setting $\ep=1$, we obtain
$$
\eta_2^* = 0.032602\; , \qquad  \eta_3^* = 0.034588\; .
$$
Thus, we conclude that the factor approximants give
$$
\eta = 0.035 \pm 0.001 \; .
$$
In the same way, we proceed with the series for $\nu^{-1}$. In the second order, 
we find
$$
A_1 = -0.869203\; , \qquad  n_1 = -0.191555\; .
$$
For the third order, we have
$$
A_1 = 1\; , \qquad n_1 = -0.152346 \; , \qquad A_2 = 0.023910\; ,
\qquad n_2 = 13.335389 \; .
$$
In the fourth order, we get
$$
A_1 = 3.027805 \; , \qquad n_1 = -0.006791 \; , \qquad A_2 = -0.440821 \; ,
\qquad n_2 = -0.424352\; .
$$
And for the fifth order, we find
$$
A_1 = 1\; , \qquad n_1 = -0.045336 \; , \qquad A_2 = 4.168053 \; ,
\qquad n_2 = -0.001772\; ,
$$
$$
A_3 = -0.312951 \; , \qquad  n_3 = -0.700494 \; .
$$
Setting $\ep=1$, for the factor approximants $\nu^*_k$, we obtain
$$
\nu^*_2 = 0.738227\; , \qquad \nu^*_3 = 0.616528\; , \qquad 
\nu^*_4 = 0.633852 \; , \qquad \nu^*_5 = 0.628417 \; .
$$
Hence, the result is
$$
\nu = 0.628 \pm 0.003 \; .
$$

Finally, reducing the series for $\om$ to form (7), we have $\om_0(\ep)=\ep$.
Following the standard procedure, for the factor approximant $\om_2^*(\ep)$, 
we get
$$
A_1 = 4.512857 \; , \qquad  n_1 = -0.139601\; .
$$
For $\om_3^*(\ep)$, we find
$$
A_1 = 1\; , \qquad n_1 = 0.006238 \; , \qquad A_2 = 4.547986 \; ,
\qquad n_2 = -0.137151\; .
$$
And for $\om_4^*(\ep)$, it follows
$$
A_1 = 4.511659 \; , \qquad n_1 = -0.139637 \; , \qquad A_2 = 107.494872 \; ,
\qquad n_2 = -0.7\times 10^{-7}\; .
$$
Setting $\ep=1$, we obtain
$$
\om^*_2 = 0.787958\; , \qquad \om^*_3 = 0.787160\; , \qquad 
\om^*_4 = 0.787934 \; .
$$
Hence, we come to the value
$$
\om = 0.788 \pm 0.0004 \; .
$$
Other critical exponents can be obtained from the scaling relations
\be
\label{9}
\al = 2 - \nu d \; , \qquad \bt =\frac{\nu}{2}\left ( d - 2 +\eta
\right ) \; , \qquad
\gm =\nu(2-\eta) \; , \qquad \dlt = \frac{d+2-\eta}{d-2+\eta} \; ,
\ee
which for the dimensionality $d=3$ simplifies to
\be
\label{10}
\al = 2 - 3\nu \; , \qquad \bt =\frac{\nu}{2}\left (1 +\eta\right )\; ,
\qquad \gm =\nu(2-\eta) \; , \qquad \dlt = \frac{5-\eta}{1+\eta} \; .
\ee
Using here the found results for the factor approximants, we have
$$
\al=0.116\pm 0.009 \; , \qquad \bt = 0.325 \pm 0.002\; ,
\qquad \gm = 1.234 \pm 0.005\; , \qquad \dlt = 4.797 \pm 0.006\; .
$$

Accomplishing in the same way calculations for the arbitrary number of 
components $N$, we obtain the factor approximants for the critical exponents 
using the general series from the Appendix A. Our results are presented in 
Table 1. It is worth emphasizing that in the two limiting cases of $N=-2$ 
and $N=\infty$, where the exact critical exponents are known, our results 
coincide with these exact values. For $N=-2$, the exact exponents are
$$
\al= \frac{1}{2}\; , \quad \bt =  \frac{1}{4}\; , \quad \gm =1 \; , 
\quad \dlt = 5 \; , \quad \eta = 0 \; , \quad \nu= \frac{1}{2}  \qquad
 (N=-2)
$$
in any dimension. And in the limit of large $N$, the exact exponents are
$$
\al= \frac{d-4}{d-2}\; , \qquad \bt =  \frac{1}{2}\; , \qquad \gm =  
\frac{2}{d-2}\; , \qquad \dlt = \frac{d+4}{d-2} \; ,
$$
$$
\eta = 0 \; , \quad \nu= \frac{1}{d-2}\; , \qquad \om= 4-d \qquad
(N\ra\infty) \; ,
$$
where $d$ is dimensionality. In three dimensions, the latter transforms to
$$
\al= -1 \; , \quad \bt =  \frac{1}{2}\; , \quad \gm =2 \; , 
\quad \dlt = 5 \; , \quad \eta = 0 \; , \quad \nu= 1\; ,  \quad
\om= 1 \; .
$$
Since our results tend to the exact values when $N\ra\infty$, the error bars 
diminish for $N\gg 1$, tending to zero, as $N\ra\infty$. Thus, for $N=100$, 
the error bar is $10^{-2}$, for $N=1000$ it is $10^{-3}$, and for $N=4$, the 
error bar is $10^{-4}$. The error bars for $N\gg 10$ diminish as $1/N$. The 
error bars for the factor approximants, up to $N=10$, obtained from the 
expansions for $\eta$, $\nu$, and $\om$, are shown in Table 2.

Critical exponents have been calculated by Monte Carlo simulations 
[18--29,34--39] and other complicated numerical methods, as is reviewed in 
Ref. [6,17,40--42]. Our results in Table 1 are in very good agreement with 
all these calculations. The advantage of our method is its simplicity. We 
have used only the expansions from the Appendix A. We do not need to know 
the large-order behavior of $\ep$-expansions, which is required for other 
methods.

\section{Exponents for spin glass}

Here we show that the method of self-similar factor approximants can be applied 
to such a notoriously difficult problem as summing the high-temperature series 
of the Ising spin glass. This model is described by the Hamiltonian
\be
\label{11}
H = -\sum_{(ij)} J_{ij} \sgm_i\sgm_j \; ,
\ee
in which $(ij)$ implies the summation over nearest neighbors, $\sgm_i$ 
takes values $\pm 1$, and $J_{ij}$ are independent random variables, whose 
dimensionless forms $\overline J_{ij}\equiv\bt J_{ij}$, with $\bt$ being
inverse temperature, occur with the probability
\be
\label{12}
p(\overline J_{ij}) = \frac{1}{2}\left [
\dlt(\overline J_{ij} - \overline J) +  
\dlt(\overline J_{ij} + \overline J) \right ]
\ee
where $\overline J_{ij}=\bt J$ is a parameter. Monte Carlo simulations 
[43--45] demonstrate the existence of the phase transition in three dimensions.

The phase transition corresponds to a singularity in susceptibilities. One 
considers two types of the latter, the Edwards-Anderson susceptibility
\be
\label{13}
\chi_{EA} \equiv \frac{1}{N}\; \sum_{i,j}  \ll \left (
<\sgm_i \sgm_j>^2 \right )\gg  \; ,
\ee
and the auxiliary susceptibility
\be
\label{14}
\chi' \equiv \frac{1}{N}\; \sum_{i,j} \left [ \ll \left (
<\sgm_i \sgm_j>^2 \right )\gg \right ]^2 \; .
\ee
Here $N$ is the total number of lattice sites, single angular brackets 
$<\ldots>$ refer to thermal averaging, and the double brackets $\ll\ldots\gg$ 
refer to averaging with respect to the distribution of interactions, defined by 
probability (12). When temperature $T$ approaches the critical temperature $T_c$, 
susceptibilities (13) and (14) behave as
\be
\label{15}
\chi_{EA} \; \propto \; (T - T_c)^{-\gm} \; , \qquad 
\chi' \; \propto \; (T - T_c)^{-\gm'} \; .
\ee

High-temperature series expansions for susceptibilities (13) and (14) are
represented as series in powers of
\be
\label{16}
w \equiv {\rm tanh}^2 (\bt J) \; .
\ee
Analyzing the series
\be
\label{17}
\chi_{EA} \simeq \sum_n a_n w^n \; , \qquad 
\chi' \simeq \sum_n a_n' w^n \; ,
\ee
one aims at finding the critical exponents $\gm$ and $\gm'$ characterizing 
the critical behavior (15). These exponents are connected with each other 
through the scaling relations
\be
\label{18}
\gm =(2-\eta)\nu \; , \qquad \gm' =(4-d -2\eta)\nu \; ,
\ee
where $\eta$ and $\nu$ are the critical exponent defining the behavior 
of the correlation function $\ll(<\sgm_i\sgm_j>^2)\gg$ and the correlation 
length $\xi\propto(T-T_c)^{-\nu}$, and where $d$ is dimensionality. Therefore, 
the exponents $\gm$ and $\gm'$ can be expressed one through another by means 
of the scaling relation
\be
\label{19}
2\gm =\gm' +\nu d \; ,
\ee
provided $\nu$ is known.

The analysis of expansions (17) turned out to be extremely difficult. This 
is because the first few terms of the series contain little information on 
spin-glass ordering. Actually, the coefficients of the first three terms 
of the series for $\chi_{EA}$ are identical to those for the susceptibility 
series of the pure Ising model. In fact, one cannot see any spin-glass 
behavior until one gets contribution from higher orders. This happens 
because an essential feature of spin glass is frustration, which reveals 
itself only in higher orders of the series. Hence, any analysis, depending 
sensitively on the first few terms in determining the critical exponents 
$\gm$ and $\gm'$, is not likely to give correct answers. These difficulties 
have been described in detail by Singh and Chakravarty [46,47], who found 
that more than ten terms in expansions (13) and (14) are necessary to be 
able to estimate the critical exponents $\gm$ and $\gm'$. They derived [46] 
in three dimensions expansions for $\chi_{EA}$ and $\chi'$ up to 17-th order
and in four dimensions, an expansion for $\chi_{EA}$ up to 15-th order.

However, even having quite a number of terms in expansions (17), it 
is very difficult to find the related critical exponents. Since, as is 
stressed above, the effects of frustration reveal themselves only in high 
orders of expansions, so that the lower orders do not provide correct 
information on spin glass behavior. The method that has been found [47] 
to be most suitable to this problem is that of inhomogeneous differential 
approximants, which is a generalization of the $d$-log Pad\'e summation. 
A weak point of this method is that the approximants, for each given order 
$k$ of an expansion, are not uniquely defined. Thus, for an expansion of 
order $k=10$, there are 42 variants of the approximants, for the expansion 
of order $k=15$, there are 96 variants, and for the order $k=17$, one has 
136 variants of different approximants. All these various approximants yield 
the results that are quite different from each other, and it is not clear 
which of them are to be accepted as correct and which as wrong, so that the 
problem arises of a subjective, not strictly defined, choice of some of them 
labelled as "well-behaved".

Now let us apply to expansions (17) the method of self-similar factor 
approximants of Section 2. In three dimensions $(d=3)$, both series for 
$\chi_{EA}$ as well as for $\chi'$ are known [47] up to seventeenth order. 
The effect of frustration, typical of spin glass, occurs in the series for 
$\chi'$ much earlier than in that for $\chi_{EA}$. Therefore more accurate 
results can be obtained considering the series for $\chi'$, which are
$$
\chi'= 1 +6w^2 + 102 w^4 - 192 w^5 + 1998 w^6 -7584 w^7 + 42822 w^8 - 
221856 w^9 +
$$
$$
+ 1147878 w^{10} - 5980608 w^{11} + 32318910 w^{12} - 167464128 w^{13} 
+ 906131742 w^{14} -
$$
\be
\label{20}
-4849958304 w^{15} + 25952889798 w^{16} - 141648771168 w^{17}\; .
\ee
For these series, we construct the factor approximants following the standard 
procedure of Section 2. The closest singularity to the origin defines the 
critical points $w_c$ and the related critical exponents $\gm'$. For the series 
of order $k=15$, we find $w_c=0.42$ and $\gm'=2.07$; for $k=16$, we have 
$w_c=0.39$ and $\gm'=1.44$; and for $k=17$, we get $w_c=0.41$ with $\gm'=1.82$. 
Thus, for the critical exponent $\gm'$, we obtain
\be
\label{21}
\gm' = 1.82 \pm 0.19 \qquad (d=3) \; .
\ee
This can be compared with the results of the Monte Carlo simulations [45], 
which find the phase transition at $T_c=1.18\pm 0.03$, with the critical 
exponent
\be
\label{22}
\gm_{MC}' = 1.87 \pm 0.28 \qquad (d=3) \; .
\ee
Using our result (21), the known value [40] of $\nu=1.3\pm 0.1$, and the 
scaling relation (19), we find the critical exponent
\be
\label{23}
\gm = 2.86 \pm 0.24 \qquad (d=3) \; .
\ee
From the Monte Carlo simulations [40] it follows
\be
\label{24}
\gm_{MC} = 2.89 \pm 0.29 \qquad (d=3) \; .
\ee
As is seen, the critical exponents in Eqs. (21) and (23) are close to the Monte 
Carlo values in Eqs. (32) and (24), respectively.

In four dimensions $(d=4)$, only the series of fifteenth order for $\chi_{EA}$
are available [47], which are
$$
\chi = 1 +8 w +56 w^2 + 392 w^3 +2408 w^4 + 15272 w^5 + 85352 w ^6 + 
508808 w^7 + 
$$
$$
+ 2625896 w^8 + 15111976 w^9 + 72067672 w^{10} +  421464680 w^{11} +
$$
\be
\label{25}
+ 1851603192 w^{12} + 11810583208 w^{13} + 46346625320 w^{14} + 
347729503368 w^{15} \; .
\ee
Constructing the factor approximants for these series, in the highest orders 
we find $w_c=0.20$, with $\gm=1.59$, for $k=14$ and $w_c=0.21$, with $\gm=2.35$,
for $k=15$. Therefore, for the critical exponent $\gm$, we obtain
$$
\gm = 2.35 \pm 0.38 \qquad (d=4)\; .
$$
To our knowledge, Monte Carlo simulations for $d=4$ are not available. And the 
method of inhomogeneous differential approximants [47] estimates $\gm\approx 
2.0\pm 0.4$. Since there is neither an expansion for $\chi'$ nor information 
on other indices, it is not possible to determine the exponent $\gm'$ in $d=4$.

The example of the present Section shows that the method of self-similar factor 
approximants can be applied to rather complicated series with very nontrivial 
behavior, requiring the consideration of high-order terms.

\section{Conclusion}

The method of self-similar factor approximants [30--33] is applied to 
calculating the critical exponents of the $N$-component vector $\vp^4$ field 
theory and of the Ising spin glass. The first example is chosen because of the 
wide interest to the $O(N)$-symmetric $\vp^4$ theory, which serves as a typical 
model for characterizing the critical behavior of a large variety of physical 
systems. We showed that the application of the method to $\ep$-expansions is 
very simple and straightforward at the same time providing the accuracy 
comparable with that of other essentially more complicated techniques.

The case of high-temperature expansions for the Ising spin glass is taken as 
an example of series with a notoriously nontrivial structure, requiring the 
consideration of high-order terms and making it very difficult an unambiguous 
determination of the exponents for susceptibilities by other known methods. Our 
method allows us to find the exponents that are in good agreement with Monte 
Carlo simulations, when the latter are available.

In the present paper, we have concentrated on the calculations of critical 
exponents. Of course, determining the critical points is also of importance. 
For instance, recently there has been a great interest to an accurate 
calculation of the critical temperature $T_c$ for interacting Bose gas (see 
review articles [48,49]). The most accurate results have been obtained so far 
by using the ideas of the optimized perturbation theory [5] in Refs. [50--55] 
and by employing Monte Carlo simulations [56--60]. These are rather involved 
numerical techniques. The method of self-similar factor approximants can also 
be applied to this problem, which, however, is a topic for a separate 
investigation.

\newpage

{\Large{\bf Appendix A}}

\vskip 5mm

The derivation of the expansions for the $N$-component field theory, considered 
in Section 2, can be found in the book [6]. These expansions are

$$
\eta(\ep)= \frac{(N + 2)\ep^2}{2(N+8)^2}
\left \{  1 + \frac{\ep}{4(N + 8)^2}\;[-N^2 + 56N + 272] -
\right.
$$
$$
-\; \frac{\ep^2}{16(N + 8)^4}\; \left [
5N^4 + 230N^3 - 1124N^2 - 17920N - 46144 + 384\zeta(3)(N + 8)(5N + 22)
\right ] - 
$$
$$
-\; \frac{\ep^3}{64(N + 8)^6}\;\left [ 13N^6 + 946N^5 + 27620N^4 + 121472
N^3 - 262528N^2 - 2912768N - 5655552 - \right.
$$
$$
- 16\zeta(3)(N + 8)\left (N^5 + 10N^4 + 1220N^3 - 1136N^2 - 68672N - 171264
\right ) + 
$$
$$
\left. \left. 
+ 1152\zeta(4)(N + 8)^3 (5N + 22) - 
5120\zeta(5)(N + 8)^2 (2N^2 + 55N + 186) \right ]\right \}  \; ,
$$
\vskip 5mm

$$
\nu^{-1} = 2 + \frac{(N+2)\ep}{N+8} \left\{ -1 -\; 
\frac{\ep}{2(N+8)^2}\;[13N+44] + \right.
$$
$$
+ \frac{\ep^2}{8(N+8)^4}\; \left [ 3N^3-452N^2-2672N - 5312 + 
96\zeta(3)(N+8)(5N+22)\right ] +
$$
$$
+ \frac{\ep^3}{8(N+8)^6}\;\left [ 3N^5+398N^4-12900N^3-81552N^2-
219968N - 357120 + \right.
$$
$$
+ 16\zeta(3)(N+8)\left (3N^4-194N^3+148N^2+9472N+19488\right ) 
+  288 \zeta(4)(N+8)^3(5N+22) - 
$$
$$
\left. - 1280\zeta(5)(N+8)^2\left (2N^2+55N+186\right )\right ] +
$$
$$
+  \frac{\ep^4}{128(N+8)^8}\; \left [3N^7 - 1198N^6 - 27484N^5 - 
1055344N^4 - 5242112N^3 - 5256704N^2 + \right.
$$
$$
+ 6999040N-626688 
- 16\zeta(3)(N+8)\left (13N^6 -310N^5 +19004N^4+ 102400N^3 - 
381536N^2 - \right.
$$
$$
- 2792576N - 4240640) 
- 1024\zeta^2(3)(N+8)^2 \left (2N^4 + 18N^3 + 981N^2 + 6994N + 11688
\right ) +
$$
$$
48\zeta(4)(N+8)^3 \left (3N^4 - 194N^3 + 148N^2 +9472N + 19488
\right ) +
$$
$$
+ 256 \zeta(5)(N+8)^2 \left ( 155N^4 + 3026N^3 + 989N^2 - 66018N - 
130608 \right ) -
$$
$$
\left. \left.
- 6400\zeta(6)(N+8)^4\left (2N^2+55N +186\right ) + 
56448\zeta(7)(N+8)^3\left (14N^2  +189N + 256\right )\right ]
 \right \} \; ,
$$
\vskip 5mm

$$
\om(\ep)= \ep - \frac{3\ep^2}{(N+8)^2}\; [3N + 14] +
$$
$$
+ \frac{\ep^3}{4(N+8)^4}\; \left [ 33N^3 + 538 N^2 + 4288N + 9568 + 
96\zeta(3)(N+8)(5N + 22)\right ] +
$$
$$
+ \frac{\ep^4}{16(N+8)^6}\; \left[ 5N^5 - 1488N^4 - 46616N^3 - 
419528N^2 - \right.
$$
$$
- 1750080N - 2599552 - 96\zeta(3)(N+8)\left (63N^3 + 548N^2 + 1916N + 
3872\right ) +
$$
$$
\left. + 288\zeta(4)(N+8)^3(5N+22) - 1920\zeta(5)(N+8)^2\left (2N^2 + 
55N + 186\right ) \right ] +
$$
$$
+ \frac{\ep^5}{64(N+8)^8}\;\left [ 13N^7 + 7196N^6 + 240328N^5 + 
3760776N^4 + \right.
$$
$$
+ 38877056N^3 + 223778048N^2 + 660389888N + 752420864 -
$$
$$
-  16\zeta(3)(N+8)\left ( 9N^6 - 1104N^5 - 11648N^4 - 243864N^3 - 
2413248N^2 - 9603328N - 14734080 \right ) -
$$
$$
-  768\zeta^2(3)(N+8)^2\left ( 6N^4 + 107N^3 + 1826N^2 + 9008N + 
8736 \right ) -
$$
$$
-  288 \zeta(4)(N+8)^3\left (63N^3 + 548N^2 + 1916N + 3872
\right ) +
$$
$$
+  256\zeta(5)(N+8)^2\left ( 305N^4 + 7386N^3 + 45654N^2 + 143212N + 
226992\right ) -
$$
$$
\left. - 9600\zeta(6)(N+8)^4\left (2N^5 +55N + 186\right ) + 
112896\zeta(7)(N+8)^3\left (14N^2 + 189N + 256\right )\right ] \; .
$$
Here $\ep=4-d$ is assumed to be asymptotically small, $\ep\ra 0$.

\newpage

\begin{center}

{\Large{\bf Table Captions}}

\end{center}

\vskip 5mm

{\bf Table 1}. Critical exponents for the $N$-component $\vp^4$ field
theory, obtained by the summation of $\ep$-expansions using the method
of self-similar factor approximants.

\vskip 5mm

{\bf Table 2}. Error bars for the critical exponent of table 1. For large 
$N\gg 10$, the error bars diminish with $N$ as $1/N$.

\newpage

\begin{center}

{\large{\bf Table 1}}

\vskip 1cm

\begin{tabular}{|c|c|c|c|c|c|c|c|} \hline
$N$&   $\al$  & $\bt$   &   $\gm$ & $\dlt$ & $\eta$  & $\nu$  & $\om$ \\ \hline
-2 &  0.5     &  0.25   &  1      & 5      & 0       & 0.5    & 0.80118 \\ 
-1 &  0.36844 &  0.27721&  1.07713& 4.88558& 0.019441& 0.54385& 0.79246\\
 0 &  0.24005 &  0.30204&  1.15587& 4.82691& 0.029706& 0.58665& 0.78832\\ 
 1 &  0.11465 &  0.32509&  1.23517& 4.79947& 0.034578& 0.62854& 0.78799\\ 
 2 & -0.00625 &  0.34653&  1.31320& 4.78962& 0.036337& 0.66875& 0.78924\\ 
 3 & -0.12063 &  0.36629&  1.38805& 4.78953& 0.036353& 0.70688& 0.79103\\ 
 4 & -0.22663 &  0.38425&  1.45813& 4.79470& 0.035430& 0.74221& 0.79296\\
 5 & -0.32290 &  0.40033&  1.52230& 4.80254& 0.034030& 0.77430& 0.79492\\ 
 6 & -0.40877 &  0.41448&  1.57982& 4.81160& 0.032418& 0.80292& 0.79694\\
 7 & -0.48420 &  0.42676&  1.63068& 4.82107& 0.030739& 0.82807& 0.79918\\
 8 & -0.54969 &  0.43730&  1.67508& 4.83049& 0.029074& 0.84990& 0.80184\\
 9 & -0.60606 &  0.44627&  1.71352& 4.83962& 0.027463& 0.86869& 0.80515\\ 
10 & -0.65432 &  0.45386&  1.74661& 4.84836& 0.025928& 0.88477& 0.80927\\
50 & -0.98766 &  0.50182&  1.98402& 4.95364& 0.007786& 0.99589& 0.93176\\
100 & -0.89650 & 0.48334&  1.92981& 4.99264& 0.001229& 0.96550& 0.97201\\
1000 & -0.99843 & 0.49933& 1.99662& 4.99859& 0.000235& 0.99843& 0.99807\\
10000 & -0.99986 & 0.49993&1.99966& 4.99986& 0.000024& 0.99984& 0.99979\\
$\infty$ &  -1   &  0.5   &  2    &  5     &  0      & 1      &  1 \\ \hline
\end{tabular}

\newpage

{\large{\bf Table 2}}

\vskip 1cm

\begin{tabular}{|c|c|c|c|} \hline
$N$ & $\eta$ error bar & $\nu$ error bar &  $\om$ error bar\\ \hline
-2  &  0     &   0     &  0.0280  \\ 
-1  &  0.0007&   0.0008&  0.0013  \\ 
 0  &  0.0010&   0.0018&  0.0110  \\ 
 1  &  0.0010&   0.0027&  0.0043  \\ 
 2  &  0.0009&   0.0034&  0.0016  \\ 
 3  &  0.0008&   0.0038&  0.0020  \\ 
 4  &  0.0007&   0.0039&  0.0016  \\
 5  &  0.0006&   0.0038&  0.0007 \\ 
 6  &  0.0005&   0.0036&  0.0005  \\
 7  &  0.0004&   0.0034&  0.0019  \\
 8  &  0.0003&   0.0032&  0.0032  \\
 9  &  0.0002&   0.0030&  0.0042  \\ 
 10 &  0.0001&   0.0029&  0.0048  \\ \hline
\end{tabular}

\end{center}
\end{document}